\documentclass[12pt,epsf,graphics]{article}
\usepackage{epsfig}
\def\be{\begin{equation}}
\def\ee{\end{equation}}
\def\bear{\begin{eqnarray}}
\def\eear{\end{eqnarray}}
\def\bearst{\begin{eqnarray*}}
\def\eearst{\end{eqnarray*}}
\begin{document}

\begin{center}
{\large\bf A model displaying extremely inhomogeneous matter distribution
  in General Relativity}
\end{center}
\vspace{1ex}
\centerline{\large Elcio Abdalla and Cecilia B. M. H. Chirenti}
\begin{center}
{Instituto de F\'\i sica, Universidade de S\~ao Paulo\\
C.P.66.318, CEP 05315-970, S\~ao Paulo, Brazil.}
\end{center}
\vspace{6ex}

\begin{abstract}

We consider a toy metric in four dimensional space-time defined in terms
of a recursive hierarchical prescription. The matter distribution turns
out to be extremely inhomogeneous. Surprisingly, for very large samples
the average mass density tends (very slowly) to a constant. There is
no trace of fractal dimension left. 

\end{abstract}

Although most authors tend to believe that the large scale distribution of
matter should be constant as predicted by Einstein's cosmological
principle \cite{cosmosprinc} there is also a belief that such an
idea should be further checked against observation
\cite{colemanpietro}. Some authors argued quite convincingly that the
large scale distribution of matter should be fractal, quite the oposite to
that expected by the cosmological principle.
Inhomogeneous cosmological models have been searched by many authors
\cite{tolmanbondi} \cite{krasinski} \cite{ribeiro} \cite{abdetal} 
\cite{abdetal2} \cite{abd}. The fact that detailed maps show quite strong
inhomogeneities has been a motivation for long controversies \cite{ribeiro2}
\cite{abdrib}.

Here, we construct a metric defined in terms of a hierarchy, building an
expectation that the matter distribution obtained from the corresponding
Einstein's Equations are extremely inhomogeneous, contradicting the
cosmological principle, and probably presenting fractality properties,
with a non trivial effective dimension. As it turns out, some of these
conclusions are false, as we shall see.

Our starting point is the diagonal metric defined in terms of a
distribution of points. A point in space-time is only defined at the
integers. Indeed, we expect that a metric describing a fractal
distribution cannot be continuous at any point, therefore it can only be
defined on a discrete set. With this motivation in mind, we define, in a
given direction $i$ (either 1, 2 or 3) the distance between two points as
pictorically represented in Fig. 1, with the metric 
\be
g_{\mu\nu}=\pmatrix{1 & 0 & 0 & 0\cr
                         0 & g_{11}&0 & 0 \cr 
                           0 & 0& g_{22}& 0 \cr  
                       0 &  0 &   0 &  g_{33} \cr   }\quad ,
\ee
where the metric is defined on all integers, depending on their
decomposition in terms of powers of 2, as below,
\bear
g_{11}(x)&=&a^{2k}\; ,\quad {\rm with}\quad x = 2^{k+1} + 2^k -1\nonumber \\
g_{22}(y)&=&a^{2l}\; ,\quad {\rm with}\quad y = 2^{l+1} + 2^l -1
\label{metriccompo}\\
g_{33}(x)&=&a^{2m}\; ,\quad {\rm with}\quad z = 2^{m+1} + 2^m -1\quad .\nonumber
\eear
Above, $k$, $l$ and $m$ are nonnegative integers and the metric is defined
as unit when the entry is an even number. This defines the metric (see
Fig. 1). 

\begin{figure*}[htb!]
\begin{center}
\leavevmode
\begin{eqnarray}
\epsfxsize= 11.truecm\rotatebox{-90}
{\epsfbox{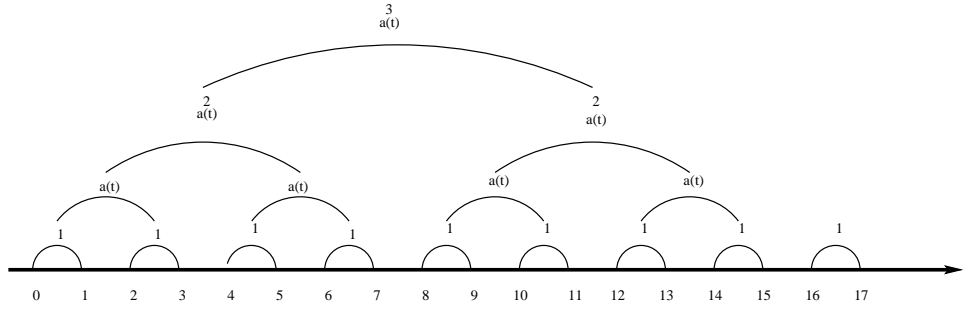}}\nonumber
\end{eqnarray}
\caption{Diagram showing the definition of a hierarchical metric as a
function of the distance. In Eq. \ref{metriccompo} we defined a metric
which realizes the above properties for nearby points. The arc segment
with a power of $a$ shown here represents the distance between any element
from one side of the arc and any element from the other side.}
\label{..}
\end{center}
\end{figure*}

Computing the Christoffel symbols and subsequently the curvature tensor
requires some care, since we are not dealing with derivatives of functions, 
but differences of functions defined on a discrete space. The result 
for the Christoffel symbols is
\bear
&&\Gamma^0_{11}=-\frac ka g_{11}\frac {da}{dt}\, ,\quad \Gamma^1_{01}=
\Gamma^1_{10}=\frac ka\frac {da}{dt}\, ,
\quad \Gamma^1_{11}=\frac{(-1)^x}2 (1- g^{11})\quad ,\nonumber\\
&&\Gamma^0_{22}=-\frac la g_{22}\frac {da}{dt}\, ,\quad \Gamma^2_{02}=
\Gamma^2_{20}=\frac la\frac {da}{dt}\, ,
\quad \Gamma^2_{22}=\frac{(-1)^y}2 (1- g^{22})\quad ,\\
&&\Gamma^0_{33}=-\frac ma g_{33}\frac {da}{dt}\, ,\quad \Gamma^3_{30}=
\Gamma^3_{03}=\frac ma \frac {da}{dt}\, ,
\quad \Gamma^3_{33}=\frac{(-1)^z}2 (1- g^{33})\quad . \nonumber
\eear
A naive construction of the Riemann tensor may lead to a result
incompatible with the general symmetry properties required for that tensor
\cite{weinberg}. We thus define it as
\be
R_{\lambda\mu\nu\kappa}=\frac 12 \lbrack 
\frac{\partial^2g_{\lambda\nu}}{\partial x^\kappa\partial x^\mu} 
-\frac{\partial^2g_{\mu\nu}}{\partial x^\kappa\partial x^\lambda} 
-\frac{\partial^2g_{\kappa\lambda}}{\partial x^\mu\partial x^\nu} 
+\frac{\partial^2g_{\mu\kappa}}{\partial x^\lambda\partial x^\nu} \rbrack
+g_{\eta\sigma} \lbrack \Gamma^\eta_{\lambda\nu} 
\Gamma^\sigma_{\mu\kappa}- \Gamma^\eta_{\lambda\kappa} \Gamma^\sigma_{\mu\nu}
\rbrack
\ee
which in the continuous version corresponds exactly to the usual
definition, and here has the correct symmetry properties with respect to
the interchange of indices. The Ricci tensor is readily obtained, with the
result (the non diagonal components vanish)
\bearst
R_{00}&=& \frac 1a (k+l+m) \frac{d^2a}{dt^2} +\frac 1{a^2}\left( 
\frac{da}{dt}\right)^2 \lbrack k(k-1) + l(l-1) + m(m-1) \rbrack \quad ,\\
R_{11}&=&  g_{11} \lbrace \frac ka +\frac 1{a^2}\left( 
\frac{da}{dt}\right)^2 \lbrack k(k-1) + kl + km \rbrack \rbrace \quad , \\
R_{22}&=&  g_{22} \lbrace \frac la + \frac 1{a^2}\left( 
\frac{da}{dt}\right)^2 \lbrack  kl + l(l-1) + lm \rbrack  \rbrace \quad ,\\
R_{33}&=& g_{33} \lbrace \frac ma +\frac 1{a^2}\left( 
\frac{da}{dt}\right)^2 \lbrack mk+ml +  m(m-1) \rbrack \rbrace \quad . \\
\eearst
The matter density can be now readily calculated by means of the Einstein
Equations, and we obtain
\bear
T_{00}\equiv \rho = \frac 1 {8\pi G} \left( 
\frac 1a \frac{da}{dt}\right)^2 (kl+lm+mk)\equiv\rho_0 (kl+lm+mk)\quad .
\eear
The result shows severe inhomogeneities. A graph showing $\rho (x,y,0)$
exemplifies the situation in Fig. 2.

\begin{figure*}[htb!]
\begin{center}
\leavevmode
\begin{eqnarray}
\epsfxsize= 12.truecm \rotatebox{-90}
{\epsfbox{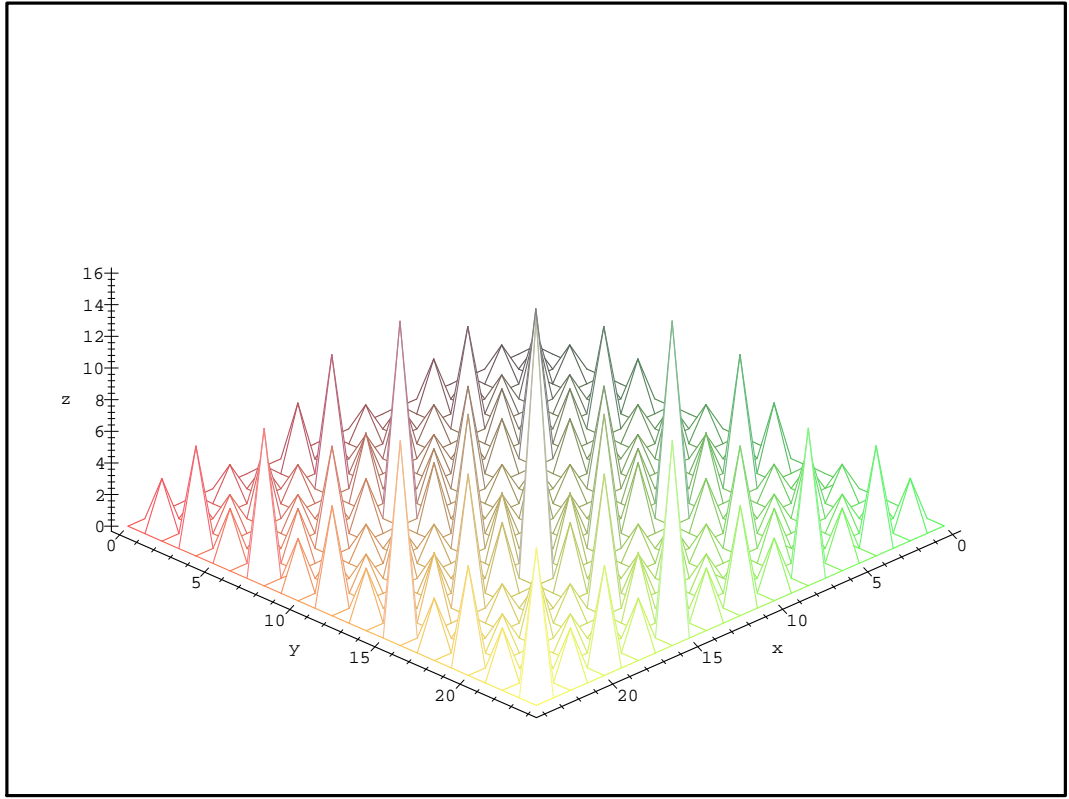}}\nonumber
\end{eqnarray}
\caption{Matter distribution for a fixed value of the $z$-component,
displaying the observed inhomogeneity.}
\label{...}
\end{center}
\end{figure*}

We further questioned about the fact of whether such an inhomogeneous
distribution could be described by a fractal. We numerically computed
the limit
\be
\lim_{r\to\infty} \frac{N(r)}{r^d}\quad ,\label{limit}
\ee
where $N(r)=\sum_{0\le x,y,z\le r} \rho (x,y,z)$ finding the value of $d$
where the limit exists. This is the usual definition of a fractal dimension
\cite{mandelbrot}. Some results are sumarized in Figure 3.

\begin{figure*}[htb!]
\begin{center}
\leavevmode
\begin{eqnarray}
\epsfxsize= 12.truecm \rotatebox{-90}
{\epsfbox{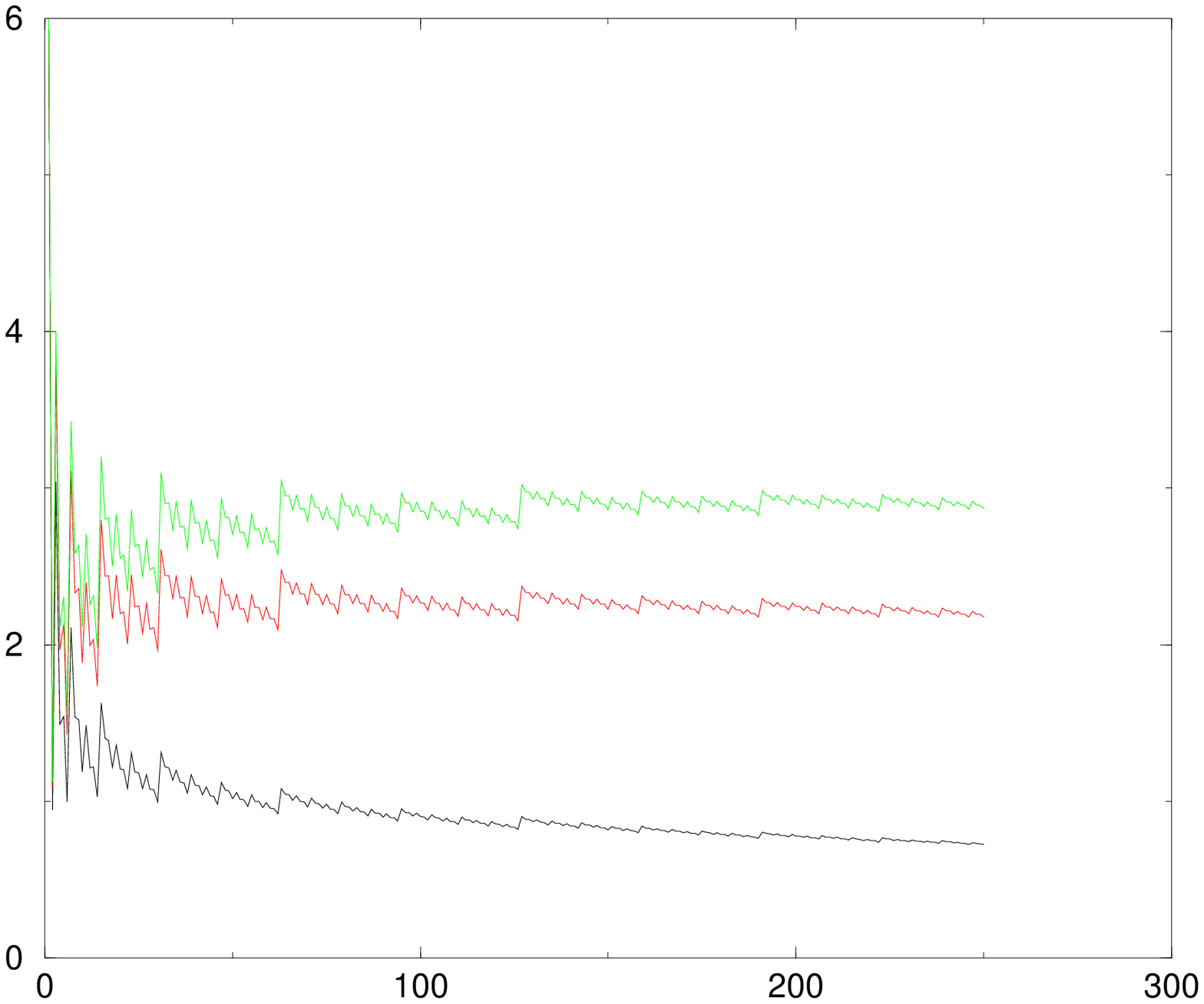}}\nonumber
\end{eqnarray}
\caption{Evolution of the average density as further data are included,
  according to the limit procedure ref{limit}. The lower diagram
  corresponds to $d=3.25$, the next one to $d=3.05$, and the highest one
  to $d=3.00$. We note that the highest one corresponds to a constant,
  thus giving the correct value of $d$. We also note the fact that we need
  a long time to achieve a cpnstant value for the limit.}
\label{....}
\end{center}
\end{figure*}

It turned out, with some surprise, that the effective dimension of the
distribution is $d=3$, that is, the hierarchical metric was not enough to
lead to a hierarchical distribution of matter. Nonetheless, the matter
distribution is very inhomogeneous, and does not seem to homogenize very
soon. Whether such a simplified model can have any relation to the real
matter distribution of the universe is not questioned, but the stability
of the dimensionality of the effective distribution of matter is quite
remarkable.  This may give support to models displaying highly
inhomogeneous behaviour either in the nonrelativistic domain, such as
\cite{abdetal}, or up to a certain cosmological distance, but smooth out
their behaviour at large enough scales, as predicted by the cosmological
principle, very slowly approaching the canonical value $d=3$ for the
effective dimension.

\bigskip
{\bf Acknowledgements:}  This work has been supported by Funda\c c\~ao
de Amparo \`a Pesquisa do Estado de S\~ao Paulo {\bf (FAPESP)} and Conselho
Nacional de Desenvolvimento Cient\'\i fico e Tecnol\'ogico {\bf (CNPq)},
Brazil.


\end{document}